\documentclass[smallextended]{svjour3}

\RequirePackage[T1]{fontenc}

\smartqed  % flush right qed marks, e.g. at end of proof

\RequirePackage{graphicx}
\RequirePackage{mathptmx}      % use Times fonts if available on your TeX system
\RequirePackage{flushend}
\RequirePackage[numbers,sort&compress]{natbib}
\RequirePackage[colorlinks,citecolor=blue,urlcolor=blue,linkcolor=blue]{hyperref}
\usepackage{amsmath}
\usepackage{amssymb}

\journalname{Gen. Rel. Grav.}

\def\to{\rightarrow}

\def\ASAS{{\it Astron. and Astrophys.} }

\def\CQG{{\it Class. Quantum Gravity} }

\def\GRG{{\it Gen. Relativ. Gravit.} }

\def\IJMP{{\it Int. J. Mod. Phys.} }

\def\JHEP{{\it JHEP} }

\def\JCAP{{\it JCAP} }

\def\MNRAS{{\it Mon. Not. R. Ast. Soc.} }

\def\NP{{\it Nucl. Phys.} }
\def\PL{{\it Phys. Lett.} }
\def\PR{{\it Phys. Rev.} }
\def\PRL{{\it Phys. Rev. Lett.} }

\def\RMP{{\it Rev. Mod. Phys.} }

\def\al{\alpha} \def\be{\beta} \def\ga{\gamma} \def\de{\delta}
\def\ep{\epsilon}   
   \def\ka{\kappa}

  \def\De{\Delta} 
   
  \def\mn{{\mu\nu}} \def\cl{{\cal L}}

 \def\frac#1#2{{\textstyle{{#1}\over
{#2}}}} 
\def\lsim{\mathrel{\rlap{\lower4pt\hbox{\hskip1pt$\sim$}}
\raise1pt\hbox{$<$}}}
\def\gsim{\mathrel{\rlap{\lower4pt\hbox{\hskip1pt$\sim$}}
\raise1pt\hbox{$>$}}} \def\sqr#1#2{{\vcenter{\vbox{\hrule height.#2pt
\hbox{\vrule width.#2pt height#1pt \kern#1pt \vrule width.#2pt} \hrule
height.#2pt}}}}
\def\square{\mathchoice\sqr66\sqr66\sqr{2.1}3\sqr{1.5}3}
\def\beq{\begin{equation}} \def\eeq{\end{equation}}
\def\beqa{\begin{eqnarray}} \def\eeqa{\end{eqnarray}}
\def\eq#1{Eq. (\ref{#1})}

\begin{document}

\title{Viability of nonminimally coupled $f(R)$ gravity}
\titlerunning{Viability of nonminimally coupled $f(R)$ gravity}

\author{Orfeu Bertolami \and Jorge P\'aramos}

\institute{Centro de F\'isica do Porto and Departamento de F\'isica e Astronomia,\\Faculdade de Ci\^encias da Universidade do Porto,\\Rua do Campo Alegre 687, 4169-007 , Porto, Portugal\\ \email{orfeu.bertolami@fc.up.pt \and jorge.paramos@fc.up.pt}}

\date{Received: \today / Accepted: }

\maketitle

\begin{abstract}
In this work we explore the viability of nonminimally coupled matter-curvature gravity theories, namely the conditions required for the absence of tachyon instabilities and ghost degrees of freedom. We contrast our finds with recent claims of a pathological behaviour of this class of models, which resorted to, in our view, an incorrect analogy with $k$-essence.
\keywords{Nonminimal Coupling \and K-Essence \and Scalar-Tensor Model}
\PACS{04.20.Fy \and 04.80.Cc \and 97.10.Cv}
\end{abstract}

\section{Introduction}

Despite its great experimental success (see {\it e.g.} Refs. \cite{Will2006,OBJP2012}), it is well known that General Relativity (GR) does not exhibit the most general form to couple matter with curvature. Indeed, these can be coupled, for instance, in a nonminimal way \cite{model} (see also Refs. \cite{model2,model3,model4} for early proposals in cosmology): this can have a bearing on the dark matter \cite{DM1,DM2} and dark energy \cite{DE1,DE2,DE4} problems (see Ref. \cite{fRDE} for a discussion in the context of $f(R)$ theories, and Ref. \cite{reviewDE} for encompassing reviews), as well as inflation \cite{preheating,Starobinsky} and structure formation \cite{Perturbations,CCS}. This putative nonminimal coupling (NMC) modifies the well-known energy conditions \cite{EC} and can give rise to several implications, from Solar System \cite{DE3} and stellar dynamics \cite{solar,Martins,collapse,spherical} to close time-like curves \cite{RZF} and wormholes \cite{Lobo}.

Following the argument that $f(R)$ theories should be derived from a more complete theory as low-energy phenomenological models \cite{Felice,SF}, one also finds strong fundamental motivation for the presence of a nonminimal coupling (NMC), as it arises from, for instance, one-loop vacuum-polarization effects in the formulation of Quantum Electrodynamics in a curved spacetime \cite{QED}, as well as in the context of multi-scalar-tensor theories, when considering matter scalar fields \cite{Damour} (as explicitly shown in Ref. \cite{scalar}). Furthermore, a NMC was put forward in an earlier  proposal \cite{Goenner}, developed in the context of Riemann-Cartan geometry, with another study showing that it clearly affects the features of the ground state \cite{ground}.

From a phenomenological standpoint, one can consider that a natural way of extending the Einstein-Hilbert action implies substituting the linear curvature term and minimal coupling with holomorphic functions $f_1(R)$ and $f_2(R)$. As such, one considers the action functional \cite{model},
\beq S = \int \left[ \ka f_1(R) + f_2(R) \mathcal{L} \right] \sqrt{-g} d^4 x~~, \label{model}\eeq
\noindent where $f_i((R)$ ($i=1,2$) are arbitrary functions of the scalar curvature, $R$, $g$ is the determinant of the metric and $\ka = c^4/16\pi G$.

By setting $f_2(R) = 1$ and $f_1(R) = f(R)$, the above encompasses the well-known $f(R)$ theories, which are widely used to study the effect of modifications of gravity in a plethora of scenarios, {\it e.g.} the Starobinsky inflationary model $f(R)=R + \al R^2$ \cite{Starobinsky}, the accelerated expansion of the Universe \cite{capoexp}, Solar System tests \cite{SS}, amongst many other studies (see Ref. \cite{Felice} for a review).

Assuming that the functions $f_i(R)$ are holomorphic, one may express them as
\beq f_i(R) = \sum_{j=-\infty}^\infty a_{ij} R^j~~, \eeq
\noindent where including negative powers of $R$ allows for the earliest $f(R) \sim R + 1/R$ dark energy models \cite{Carroll}. Inverse powers of $R$ clearly require that the background cosmological value of the scalar curvature is non-vanishing, as is characteristic of {\it e.g.} an exponential or power-law evolution of the scale factor.

The description of the non-trivial functions $f_i(R)$ as a combination of infinite terms allows in principle the probing of individual terms of the summation by ascertaining the dynamical impact of the action Eq. (\ref{model}) in a specific setting where a particular term $R^j$ is dominant --- from the astrophysical \cite{solar} and Solar System \cite{DE3} to galactic \cite{DM1} and cluster \cite{DM2} scenarios, up to a cosmological scale \cite{DE1,DE2,DE4};  this justifies the assumption of simple power laws for the generic functions $f_i(R) \sim R^{n_i}$.

Interestingly, a recent study \cite{Ribeiro} has shown that the choice $f_1(R) \sim R^{1+n}$ and $f_2(R) \sim R^n$ appears to negate the cosmological impact of the additional dynamics found in Eq. (\ref{field0}); indeed, this choice enables the rewriting of the action functional Eq. (\ref{model}) as
\beq S = \int \left[ \ka R + \mathcal{L} \right] \left({R\over M}\right)^{n} \sqrt{-g} d^4 x~~, \label{model2}\eeq
\noindent where $M$ is a characteristic mass scale to be determined. This shows that such a subset of models can be viewed in terms of a generalised measure replacing the usual invariant $\sqrt{-g}d^4x$.

Variation with respect to the metric yields the modified field equations,
\beqa \label{field0} \left( \ka F_1 + F_2 \cl \right) G_\mn &=&  {1 \over 2} f_2 T_\mn + \De_\mn \left( \ka F_1 + F_2 \cl \right) \\ \nonumber && + {1 \over 2} g_\mn \left[ \ka(f_1 - F_1 R) - F_2 R \cl \right] ~~, \eeqa
\noindent with $F_i \equiv df_i/dR$ and $\De_\mn = \nabla_\mu \nabla_\nu - g_\mn \square$. As expected, GR is recovered by setting $f_1(R) = R $ and $f_2(R) = 1$, while the usual field equations of $f(R)$ theories are obtained from $f_2(R) =1$ \cite{Felice}. 

The trace of \eq{field0} reads
\beq \left( \ka F_1  + F_2  \cl \right) R  =   {1 \over 2} f_2 T -3 \square \left( \ka F_1 + F_2 \cl \right) + 2 \ka f_1 ~~. \label{trace0}\eeq
Resorting to the Bianchi identities, one concludes that the energy-momentum tensor of matter may not be (covariantly) conserved, since
\beq \nabla_\mu T^\mn={F_2 \over f_2}\left(g^\mn \cl-T^\mn\right)\nabla_\mu R~~, \label{cov} \eeq
\noindent can be non-vanishing (see Refs. \cite{KoivistoCov,Sotiriou1} for thorough discussions).

\section{Equivalence with a two-scalar tensor theory}

In Ref. \cite{scalar} (see also Refs. \cite{equivalence2}), it was shown that \eq{model} is equivalent to the action involving two additional scalar fields $\phi, \psi$ (with dimensions of mass-squared),
\beq S = \int \left[ \psi (R-\phi) + \ka f_1(\phi) + f_2(\phi) \mathcal{L} \right] \sqrt{-g} d^4 x~~. \label{model3}\eeq
Indeed, the variation of the above with respect to the scalar fields leads to
\beq \phi = R ~~~~,~~~~\psi = \ka F_1(\phi) + F_2(\phi)\cl ~~,\eeq
\noindent which, upon replacing in Eq. (\ref{model3}), yields the original action functional Eq. (\ref{model}).

The above action is in the Jordan frame, as the scalar field $\psi$ appears coupled to the scalar curvature $R$. One may perform the conformal transformation (while maintaining the original coordinate system) $ \tilde{g}_\mn = \psi g_\mn$, so that 
\beq \sqrt{-g} = \psi^{-2} \sqrt{-\tilde{g}}~~~~,~~~~R = \psi \left[ \tilde{R} - 6 \sqrt{\psi} \tilde{\square} \left({1 \over \sqrt{\psi}} \right) \right] ~~,\eeq
\noindent where $\tilde{\square}$ denotes the D'Alembertian operator constructed from the metric $\tilde{g}_\mn$.

Introducing the above relations into Eq. (\ref{model3}) leads to a decoupling of the rewritten scalar curvature $\tilde{R}$ from the scalar field $\psi$, so that the action functional is now in the Einstein frame:
\beq \label{equivaction} S = \int  \sqrt{-g} d^4x \bigg( 2 \ka \left[ R - 2g^\mn \varphi^1_{,\mu} \varphi^1_{,\nu} -  4 U(\varphi^1,\varphi^2) \right] + {\cal L'} \bigg)~~, \eeq
\noindent where $\varphi^1$ and $\varphi^2$ are two scalar fields, related with the scalar curvature and the non-trivial $f_1(R)$ and $f_2(R)$ functions through
\beq \label{deffields} \varphi^1 = {\sqrt{3}\over 2} \log \left[ F_1(R) + {F_2(R) {\cal L} \over 2\ka} \right] ~~~~,~~~~ \varphi^2 = R ~~, \eeq
\noindent where the potential is given by 
\beq \label{potential} U(\varphi^1,\varphi^2) = {1 \over 4} \exp \left( -{2 \sqrt{3}\over 3} \varphi^1 \right) \left[\varphi^2 - f_1(\varphi^2 ) \exp \left( -{2 \sqrt{3}\over 3} \varphi^1 \right)  \right]~~, \eeq
\noindent and a modified Lagrangian density is used when constructing the Lagrangian density of the latter,
\beq \label{clstar} {\cal L' } (\varphi^1,\varphi^2,g_\mn, \psi)= \exp [-(4\sqrt{3}/3)\varphi^1] f_2(\varphi^2) \cl(g'_\mn, \psi) ~~, \eeq
\noindent where $\psi$ denotes existing matter fields. Notice that the non-minimal coupling present in the definition above is retained throughout the derivation, and that both $f_1(R) $ and $f_2(R)$ add an additional coupling through both the exponential term depending on $\varphi^1$ and the ``physical metric''
\beq \label{physicalmetric} g'_\mn = \exp [-(2\sqrt{3}/3)\varphi^1] g_\mn~~,\eeq
\noindent which is used to construct the original Lagrangian density $\cl$ of matter ({\it e.g.} through the contraction of indices).

The above can be better understood through the use of the specific example: for this, one adds next-to-leading order terms for both $f_1(R) $ and $f_2(R)$, as previously explored in Ref. \cite{preheating}: 
\beq f_1(R) = R + {R^2 \over 6M^2} ~~~~,~~~~f_2(R) = 1 + 2\xi{R \over M^2}~~,\eeq
\noindent where $M \sim 10^6 M_{P}$ is the characteristic mass scale required for Starobinsky inflation \cite{Starobinsky}. This is equivalent to the formulation Eq. (\ref{equivaction}) in the Einstein frame, with the scalar fields
\beq \varphi^1 = {\sqrt{3}\over2} \log \left( 1 + {R \over 3M^2} + 2\xi { {\cal L} \over \ka M^2 } \right) ~~~~,~~~~ \varphi^2 = R ~~, \eeq
\noindent driven by the potential
\beq U(\varphi^1,\varphi^2) =   {1 \over 4} \varphi^2 \exp \left( -{2 \sqrt{3}\over 3} \varphi^1 \right) \left[1 - 
\left(1 + {\varphi^2 \over 6M^2} \right)  \exp \left( -{2 \sqrt{3}\over 3} \varphi^1 \right)  \right]~~. \eeq

\section{Gravity propagator of a NMC model}

In this section one approaches the issue of directly computing the gravity propagator of the NMC model under consideration. The improper use of the equivalence with a multi-scalar-tensor theory \cite{scalar} has led to misinterpretations reported in Ref. \cite{Koivisto} and discussed in the following section. For robustness, the computation of the propagator is discussed without resorting to the aforementioned equivalence multi-scalar-tensor theory (see Ref. \cite{fRT} for an analysis of alternative $f(R,R_\mn,T_\mn,T)$ theories).

The (inverse) gravity propagator around Minkowski spacetime is given by

\beq \Pi_{\mu\nu}^{-1}{}^{\lambda\sigma}h_{\lambda\sigma}={1 \over 2\ka} \tau_{\mu\nu}~~, \label{invprop} \eeq

\noindent where $\tau_\mn $ is the energy-momentum tensor of matter and $h_\mn $ is a small perturbation to the Minkowski background metric $\eta_\mn$, {\it i.e.} $g_\mn = \eta_\mn + h_\mn$.
 
To compute it, the (4-rank tensor) spin projector operators are introduced \cite{prop},

\begin{eqnarray}
{\cal P}^2&=&{1 \over 2}(\theta_{\mu\rho}\theta_{\nu\sigma}+\theta_{\mu\sigma}\theta_{\nu\rho})-{1 \over 3}\theta_{\mu\nu}\theta_{\rho\sigma}~~,\\ \nonumber
{\cal P}^1&=&{1 \over 2}(\theta_{\mu\rho}\omega_{\nu\sigma}+\theta_{\mu\sigma}\omega_{\nu\rho}+\theta_{\nu\rho}\omega_{\mu\sigma}+\theta_{\nu\sigma}\omega_{\mu\rho})~~,\\ \nonumber {\cal P}^0_s&=&{1 \over 3}\theta_{\mu\nu}\theta_{\rho\sigma}~~,~~
{\cal P}^0_w=\omega_{\mu\nu}\omega_{\rho\sigma}~~,~~
{\cal P}^0_{sw}={1 \over \sqrt{3}}\theta_{\mu\nu}\omega_{\rho\sigma}\,, \quad
{\cal P}^0_{ws}={1 \over \sqrt{3}}\omega_{\mu\nu}\theta_{\rho\sigma}\,,
\end{eqnarray}
with the transversal and longitudinal projectors in momentum space given by
\begin{equation}
\theta_{\mu\nu}=\eta_{\mu\nu}-{k_\mu k_\nu \over k^2}~~,~~ \omega_{\mu\nu}={k_\mu k_\nu \over k^2}~~.
\nonumber
\end{equation}
 
\noindent The set $\{{\cal P}^2,{\cal P}^1,{\cal P}_s^0,{\cal P}_w^0\}$ constitute a complete set of projection operators
\beq
{\cal P}^{i}_{a}{\cal P}^{j}_{b}=\delta^{ij}\delta_{ab}{\cal P}^{i}_{a}~~,~~{\cal P}^2+{\cal P}^1+{\cal P}_s^0+{\cal P}_w^0=1~~,
\eeq
as one can easily verify.

${\cal P}^2$ and ${\cal P}^1$ represent the four degrees of freedom of transverse and traceless spin-2 and spin-1 degrees, while ${\cal P}_s^0$, ${\cal P}_w^0$ represent spin-0 scalar multiplets. Furthermore, ${\cal P}_{sw}^0$ and ${\cal P}_{ws}^0$ mix the two scalar multiplets, according to
\beq{\cal P}^0_{ij}{\cal P}^0_k=\delta_{jk}{\cal P}^0_{ij}~~, ~~ {\cal P}^0_{ij}{\cal P}^0_{kl} = \delta_{il}\delta_{jk}{\cal P}^0_k~~, ~~ {\cal P}^0_k {\cal P}^0_{ij} = \delta_{ik}{\cal P}^0_{ij}~~.
\eeq

As shown in Refs. \cite{propfR}, the gravity propagator of $f(R) $ theories in a flat spacetime is given by
\beq \Pi_f = \Pi_{GR} + {1 \over 2}{{\cal P}^0_s \over k^2+m^2}~~,\eeq
\noindent where $m^2 \equiv (3f''(0))^{-1} $ and 
\beq \Pi_{GR} = {1 \over k^2} \left( {\cal P}^2 -{1 \over 2} {\cal P}^0_s\right)~~,\eeq
\noindent is the gravity propagator of GR. The above clearly depicts the additional scalar degree of freedom introduced by $f(R)$ theories, as known from the equivalence between the latter and a scalar-tensor theory \cite{equivalence2}.

In order to obtain the above propagator, the modified field Eqs. (\ref{field0}) (with $f_2(R) = 1$) were expanded around a flat spacetime, together with the leading order expansion $ f(R) = f_1(R) = 2\Lambda + R + R^2 /(6m^2)$. To extend this result to a non-trivial NMC $f_2(R) \neq 1$, it suffices to perform the same procedure to the additional terms introduced by the latter, assuming the leading-order expansion below:

\beq f_2 (R) = 1 + {R \over 6M^2} \to F_2 = {1 \over 6M^2}~~.\eeq

Notice that one could alternatively redefine the energy-momentum tensor $\tau_\mn$ as the variation of the NMC Lagrangian density term,

\beq \tau_\mn = -{2 \over \sqrt{-g}} {\de \left[\sqrt{-g} f_2(R)\mathcal{L}_m \right] \over \de g^\mn } = f_2 T_\mn + 2\De_\mn \left( F_2 \cl \right) - 2 F_2 \cl R_\mn~~, \eeq

\noindent where $T_\mn$ is the usually defined energy-momentum tensor of matter.

Strikingly, the above shows that the linearized field equations cannot be written in the form \eq{invprop}, as the matter Lagrangian density $\cl$ and the metric perturbation and its derivatives $h_\mn, h_{\mn,\al\be}$ appear coupled: instead, applying the spin projector formalism would eventually yield an irreducible equation of the form

\beq {(\Pi_{f}^{-1})}_{\mu\nu}^{\lambda\sigma}h_{\lambda\sigma}={1 \over 2\kappa} g[{({\cal P}_i^j)}^{\alpha\beta}_{\mu\nu} h_{\alpha\beta} \cl,T_{\mu\nu}]~~, \eeq

\noindent which, even if one could write the Lagrangian density $\cl$ as a function of the energy-momentum tensor $T_\mn$ and the metric $g_\mn = \eta_\mn + h_\mn$, would still possess cross-terms between $T_\mn $ and $h_{\al\be}$. As such, one finds that the direct derivation of a gravity propagator is unfeasible, since the NMC directly implies that one cannot write the metric perturbation as given by the application of spin projectors to the energy-momentum tensor.

\section{Analogy with k-essence model}

In a recent study \cite{Koivisto}, Tamanini and Koivisto claim that a NMC model is not viable, due to the appearance of ghost degrees of freedom. To derive this, Ref. \cite{Koivisto} starts by integrating out the auxiliary scalar field $\varphi^2$, thus rewriting the action, \eq{equivaction}, as
\beq \label{Koivistoaction} S = \int  \sqrt{-g} d^4x \left[  R - {1 \over 2} g^\mn \varphi^1_{,\mu} \varphi^1_{,\nu} - p(\cl, \varphi^1) \right]~~, \eeq
\noindent where $p(\cl,\varphi^1)$ is a function obtained from the substitution of the solution $\varphi^2 = \varphi^{2*}$ of the field equation for this scalar field.

The authors then argue that a viable NMC model should allow the coupling to any Lagrangian density for matter $\cl$ , and choose
\beq
\cl = - {1\over 2}g^\mn \chi_{,\mu}\chi_{,\nu} \equiv X~~,
\eeq
\noindent {\it i.e.} a purely kinetic scalar field with no potential. The above action with this choice is then compared with purely kinetic k-essence models \cite{kessence1,kessence2}, given by the action 
\beq \label{kessenceaction} S =  \int  \sqrt{-g} d^4x  \left[ 2 \ka R + p(X) \right] ~~.\eeq
This is done with little regard for the differences between k-essence and a NMC model, as no in depth justification is offered: instead, it is simply stated that such comparison is warranted because the kinetic term of $\varphi^1$ is canonical, so that its contribution (both the canonical kinetic term and the effect on $p(X,\varphi^1)$ can be disregarded.
Conditions for positivity of the energy density $\ep(\chi)$ of the matter scalar field $\chi$ and the speed of sound $c_s(\chi)$ at which its perturbations propagate are then introduced \cite{speed1,fluid,kessence2},
\beq \label{kenergy} \ep(\chi) = 2 X_0 p_{,X}(X_0) - p(X_0) >0~~, \eeq
\noindent and
\beq \label{kspeed} c_s^2(\chi) = {p_{,X}(X_0) \over 2X_0 p_{,X X}(X_0) + p_{,X}(X_0)}\geq 0~~,\eeq
\noindent  and used to argue severe constraints on the form of the NMC function $f_2(R)$. Notice that the above expression is evaluated at the background value $X= X_0$ \cite{kessence2}. Thus, the case $c_s = 0$ does not necessarily imply that the pressure is independent of the kinetic term $X$ defined above (as would be the case for a Cosmological Constant, where the associated pressure and density are constant): it may simply happen that the evaluation of a complex form at the background value vanishes, {\it i.e.} $p_{,X}(X) \neq 0$ but $p_{,X}(X_0) =0$ \cite{speedvanish}.

Inspection shows that the comparison between \eq{Koivistoaction} and (\ref{kessenceaction}) leaves the dynamical effect of the scalar field $\varphi^1$ out of the ensuing discussion: this additional degree of freedom should have some impact on what otherwise bears a resemblance with k-essence models, specially since it is not an independent matter field, but is dynamically related to both the curvature and the Lagrangian density of matter through \eq{deffields}.

As such, the purpose of this work is not to claim that no pathological behaviour can arise in NMC models --- as indeed one knows that the condition $f'(R) >0$ is required for the absence of ghosts in $f(R)$ theories \cite{Starobinsky_ghosts}, while in NMC models it is known that Dolgov-Kawasaki instabilities appear if $F_1' + F_2' \cl $ is negative \cite{EC,DK,OdintsovDK,FaraoniDK}: notwithstanding the need for future work, this comment aims at showing that the recent study by Tamanini and Koivisto does not address the problem at hand in a rigorous fashion. Several remarks are in order, as discussed below.
 
\subsection{Jordan versus Einstein frame}
 
As pointed out following \eq{clstar}, the NMC manifests itself in \eq{equivaction} by both the original NMC (now as a function of $\varphi^2$), a factor arising from the conformal transformation from the original Jordan frame of \eq{model} to the Einstein frame, and the physical metric resulting from the latter. In particular, Ref. \cite{Perturbations} has shown that modifications to cosmological perturbations arise both from the effect of the modified field Eqs. (\ref{field0}) as well as the non-conservation of the energy-momentum tensor, \eq{cov} --- something which is not captured in the direct comparison with k-essence.

Thus, if one aims at studying a NMC purely kinetic scalar field, $\cl =X$ cannot be introduced directly in \eq{Koivistoaction}, as one should recall that the kinetic term of $\chi$ cannot be constructed from the metric $g_\mn$, but instead from the physical metric \eq{physicalmetric}, so that one obtains
\beq \label{diffconf} X' = {1 \over 2} g'^\mn \chi_{,\mu} \chi_{,\nu} = {1 \over 2} \exp \left({2\sqrt{3}\over 3}\varphi^1\right) g^\mn \chi_{,\mu} \chi_{,\nu} = \exp \left({2\sqrt{3}\over 3}\varphi^1\right)X ~~, \eeq
\noindent where one recalls that the solution $\varphi^2 = \varphi^{2*}$ depends on both $X $ and $\varphi^1$.

Thus the k-essence function is of the form $p=p(X',\varphi^1)=p(\exp [(2\sqrt{3}/3)\varphi^1]X ,\varphi^1)$, with a dependence on $\varphi^1$ that is much more evolved than assumed in Ref. \cite{Koivisto} --- which implies that neglecting these terms is improper.

\subsection{Incorrect background comparison}

Conditions \eq{kenergy} and (\ref{kspeed}) are based upon the study of perturbations of the k-essence field $\chi$ around the cosmological FRW background, so that the kinetic term $X$ is evaluated at its background cosmological value \cite{kessence2} --- which is therefore constrained to be positive, $X>0$, and cannot assume any arbitrary value, as also claimed in Ref. \cite{Koivisto}.

As discussed in the previous paragraph, the correct kinetic term should be computed using the physical metric $g'^\mn$, leading to \eq{diffconf}. Since the derivation of \eq{kenergy} and (\ref{kspeed}) in the context of k-essence has no such distinction, the underlying computation of perturbations cannot be applied to the present context. 

Furthermore, the calculations leading to \eq{kenergy} and (\ref{kspeed}) assume that the cosmological evolution is driven by k-essence (the dominant contribution to the dynamics, which also include minimally coupled baryonic matter) \cite{kessence2}, so that the analysis of Ref. \cite{Koivisto}, if correct, would only apply to a proposal for dark energy based upon a NMC scalar field with no potential term: as such, it would only be useful in ruling out such a model --- but not the existing proposal for a phase of accelerated expansion of the universe due to a NMC between curvature and baryonic matter \cite{DE1,DE2,DE4}, which is described by $\cl = -\rho$ \cite{Lobo}, its energy density, and requires no additional scalar field.

\subsubsection{Speed of sound of curvature perturbations}

Given the above, it is clear that a correct treatment of cosmological perturbations of a NMC purely kinetic scalar field requires the inclusion of the dynamical contribution of the additional degree of freedom embodied in $\varphi^1$. Indeed, considering the perturbed metric 
\beq ds^2 = (1+2\Phi)dt^2 - (1-2\Psi) a^2(t) \ga_{ij} dx^i dx^j ~~. \eeq
\noindent and assuming small perturbations $\chi = \chi_0 (t) + \de \chi (t,\vec{x})$, $\varphi = \varphi_0 (t) + \de \varphi (t,\vec{x})$ yields, to first order in all perturbations, the field equation for $\chi$ \cite{private},
\beqa && 0 = P_{,X} {k^2 \over a^2} \de \chi + P_{,X \varphi} \dot{\chi}_0 \de \dot{\varphi} - 3 P_{,X} \dot{\chi}_0 \dot{\Psi} + \\ \nonumber && (P_{,X} + P_{,XX} \dot{\chi}_0^2 ) ( \de \ddot{\chi} + 3 H \de \dot{\chi} - \dot{\chi}_0 \dot{\Phi}) + \\ \nonumber && [ ( 3 P_{,XX} + P_{,XXX} \dot{\chi}_0^2 ) \dot{\chi}_0 \ddot{\chi}_0 + (P_{,X\varphi} + P_{,XX\varphi} \dot{\chi}_0) \dot{\varphi}_0^2 ] \de \dot{\chi} +  \\ \nonumber && [(P_{,X \varphi} + P_{,XX\varphi} \dot{\chi}^2_0 ) \ddot{\chi}_0 + (3H P_{,X\varphi} + P_{,X \varphi\varphi} \dot{\varphi}_0 ) \dot{\chi}_0 ] \de \varphi \\ \nonumber && + [(3H P_{,XX} + P_{,X \varphi} \dot{\varphi}_0 ) \dot{\chi}_0^3 - 2 (3H P_{,X} + P_{,X\varphi} \dot{\varphi}_0) \dot{\chi}_0 - \\ \nonumber && (2 P_{,X} + 5 P_{,XX} \dot{\chi}_0^2 + P_{,XXX} \dot{\chi}_0^4 ) \ddot{\chi}_0] \Phi ~~.  \eeqa
\noindent If one simply takes the ratio between the second time derivative and the gradient terms, \eq{kspeed} ensues. As shown below, inclusion of the additional degree of freedom $\varphi^1$ yields different results, showing that the above computation ignores the effect of other terms (such as the gravitational potentials $\Phi$ and $\Psi$), which are related to $\de \varphi^1$ through the perturbed Einstein equations.

The authors of this work do not know of any study of cosmological perturbations of a interacting two-field k-essence model, precisely due to the intricate nature of the ensuing equations of motion (and its perturbations); existing works on interacting \cite{multifield1,multifield1b}, non-interacting \cite{multifield2} and purely kinetic multifield k-essence models \cite{multifield3} do show that the relevant criteria for model stability is that the speed of sound of curvature perturbations should be positive defined, and that the latter is related to the speed of sound of all the scalar fields considered.

Notice that, since the scalar field $\varphi^1$ has a canonical kinetic term, one cannot use the matrix formalism of Ref. \cite{multifield1} to assess the propagation of curvature perturbations in an interacting multifield k-essence model, which is only valid for non-linear kinetic terms.

For the sake of argument we study what would happen if the contribution of $\varphi^1$ to the k-essence function could somehow be disregarded (as assumed in Ref. \cite{Koivisto}) and forego this interaction term of \eq{Koivistoaction}, $p(X,\varphi^1) \approx p(X)$, but keep the kinetic term of latter --- thus considering instead the action
 
\beq \label{multifieldaction} S = \int  \sqrt{-g} d^4x \left[ 2\ka R - P(X,Y) \right]~~, \eeq

\noindent where $P(X,Y) = P_X(X) + P_Y(Y) \equiv p(x) + Y$ and $2Y = - g^\mn \varphi^1_{,\mu} \varphi^1_{,\nu} $. One may now resort to Ref. \cite{multifield2}, where it was shown that the speed of sound of curvature perturbations is given by
\beq c_s^2 = \sum (\rho_i + P_i) \left( \sum {\rho_i + P_i \over c_i^2 } \right)^{-1} ~~~~,~~~~i=X,Y~~.\eeq
Using Eqs. (\ref{kenergy}) and (\ref{kspeed}) for the contribution of $X$ and $\rho_Y = P_Y = Y$, so that $c_Y = 1$ (echoing the well-known result that perturbations of a canonical scalar field propagate at the speed of light), one gets 

\beq c_s^2 = { X p_{,X} + Y \over  {X ( 2X p_{,X X} + p_{,X}}) + Y} ~~,\eeq

\noindent which collapses into \eq{kspeed} when $Y=0$. This highlights the relevance of the contribution of $\varphi^1$, and shows that the analysis of Ref. \cite{Koivisto} is only correct if the latter can be disregarded, as argued before. Again, evaluating this expression would require a knowledge of both $X$ and $Y$ derived from a cosmological model for dark energy arising from a NMC purely kinetic scalar field, which has nothing in common with the proposal put forward in Ref. \cite{DE1,DE2,DE4}, and as such is not pursued: instead, the aim of the above discussion is to show the incorrect use of an analogy with a single scalar field pure kinetic k-essence model.

\subsection{Causality}

Finally, the constraints on NMC models obtained in Ref. \cite{Koivisto} are derived from the imposition that the speed of sound is lesser than that of light, $c_s \leq c$, due to the usual argument for causality. However, following Ref. \cite{Ellis}, one may instead defend that obtaining a faster-than-light propagation of perturbations for a particular matter species does not invalidate the underlying theory, but instead shows that the adopted matter content is unphysical (in the context of the particular model under scrutiny).

Thus, the conclusions of Ref. \cite{Koivisto} refer to the rather unnatural assumption of NMC k-essence and does not have any bearing on models for dark matter based on NMC baryonic matter.

Furthermore, this prohibition of a superluminal speed of sound is not as obvious as it may appear at first sight: indeed, Lorentz invariant theories are allowed with $c_s > c$, since the speed of sound can then act as a causality-imposing upper limit for the velocity --- albeit at the cost of possibly ill-defined microscopic properties of matter, and with the caveat that $c_s$ is then fundamentally unbounded itself \cite{Ellis}.

A more thorough analysis of the Lorentzian signature of the spacetime manifold relies on global hyperbolicity as a sufficient requirement for causality. While this translates into the condition $c_s < c$ in a static spacetime, non-trivial backgrounds can suitably evade this inequality, thus rendering the constraints obtained in Ref. \cite{Koivisto} devoid of meaning (see Ref. \cite{Babichev} for an extended discussion in the context of k-essence and Ref. \cite{Bruneton} for a survey of the issue in alternative gravity models).

\section{Discussion and Outlook}

In this work we discuss the viability of NMC models, and argue that the recent claims of pathological behaviour of non-minimally coupled models are not conclusive, due to the improper comparison between the latter and k-essence: our argument is based upon four major, independent points:

1. The use of results from single scalar field purely kinetic k-essence implies that the background cosmological dynamics (namely the current accelerated expansion of the universe, {\it i.e.} dark energy) is described by the latter or, as Ref. \cite{Koivisto} argues, from a NMC scalar field with $\cl = X$: as such scenario has never been put forward (and is highly unnatural to begin with), but the existing proposals rely instead on NMC baryonic matter (a much more plausible assumption) \cite{DE1,DE2,DE4}, the comparison with k-essence essentially does not refer to the considered models.

2. The conformal transformation used to arrive at the equivalent action \eq{equivaction} relies on a conformal transformation from the Jordan to the Einstein frame, which implies that the Lagrangian density of matter should be constructed with a physical metric $g'_\mn$ related to $\varphi^1$: this enables a more complex dependency of the putative k-essence function $p$ on $\varphi^1$, which should not be dismissed.

3. Even if this interaction between the purely kinetic Lagrangian density $\cl = X $ and the additional degree of freedom $\varphi^2$ could somehow be disregarded, the effect of the canonical term of the latter leads to a speed of sound for curvature perturbations different from the single field k-essence form argued in \cite{Koivisto} --- again highlighting the relevance of $\varphi^2$.

4. Finally, the use of the constraint $c_s^2 < 1$ is not completely clear or universally accepted, as the structure of spacetime in an evolving universe implies that the usual causality argument established in Minkowski spacetime may no longer be valid.

Given the above, we believe that the analysis of Tamanini and Koivisto \cite{Koivisto} is unfounded, as it presents an incorrect analogy with k-essence which, even if correct, would not apply to models in which there is only baryonic matter. We naturally concede that the interesting questions raised in that work prompt further investigation --- namely how the analysis of the propagation of perturbations in NMC cosmological models can lead to further constraints on the model, and how an NMC is related to causality and the global structure of spacetime.

\begin{acknowledgements}
The authors would like to thank F. S. N. Lobo for fruitful discussions, and the referee for his/her valuable remarks and criticism. The work of O.B. and J.P. was partially supported by FCT (Funda\c{c}\~ao para a Ci\^encia e a Tecnologia, Portugal) under the project PTDC/FIS/111362/2009. 
\end{acknowledgements}

\end{document}